\documentclass[twocolumn,showpacs,preprintnumbers,pra,amsmath,amssymb]{revtex4}

\usepackage{graphicx}
\usepackage{bm}
\begin{document}
\title{An Instructional Scaffolding for Intuitive Explanation of ``Why does not a spinning top collapse?"}

\author{Won-Young HWANG$^*$}

\affiliation{ Department of Physics Education, Chonnam National
University, Gwangju 500-757, Republic of Korea}

\begin{abstract}
``Why does not a spinning top collapse?" is a puzzling question. Standard solution using angular momentum and torque is not intuitive enough. Thus intuitive explanations for the question have been proposed. We provide scaffolding for an intuitive explanation for the question. Accelerated point-masses in the top exert forces on the frame, which balances the effect due to gravity. The explanation is supplemented by the two following points. A more rigorous conceptual framework of the explanation is provided. A full calculation of trajectory is given. Nutation of spinning top is a difficult issue to understand physically. However, the nutation can also be understood by the intuitive explanation. We discuss another intuitive explanation.
\end{abstract}
\maketitle
\section{Introduction}\label{sec:intro}
``Why does not a spinning top collapse?" is a question that has puzzled many physicists. A standard solution can be found in
textbooks, e.g. Refs. \cite{Hal93,Fey63,Tho95}: Torque on a top is
given such that angular momentum of the top makes a closed loop
around the vertical axis. The solution is simple and rigorous.
However, the solution is not intuitive enough. This situation is not so satisfactory. Students beginning to study physics often experience difficulties in understanding true meaning of explanations.

Intuitive explanations were given by Barker \cite{Bar60}, Eastman \cite{Eas75}, and Edwards \cite{Edw77}. Explanation by Eastman and Edwards is interesting but somewhat confusing. Explanation by Barker got to the heart of the problem; Accelerated point-masses in the top exert forces on frame to generate torque which balances torque due to gravity. However, there are two points to be supplemented; Conceptual framework of the explanation is not presented in detail.
Although idea for the calculation is provided with correct results, full trajectory is not given. These points will make the explanation more difficult to be understood. The purpose of this paper is to supplement the explanation in the points. Our presentation may work as a scaffolding for the explanation by Barker.

A rotating object is simpler than the spinning top. But both objects have the same nature with respect to its explanations. Thus we adopt the rotating object as preparation step in section II-A. In section II-B, we give the explanation for the spinning top. Core of the explanation can be found in Fig. 2 with its caption. Section II-C deal with nutation of spinning top. Eastman's explanation is discussed in Section III. Then we summarize in the last section.

\section{Scaffoldings for Intuitive Explanation}\label{sec:II}

\subsection{Step 1: A rotating object}
\begin{figure}[htbp]
\includegraphics[width=7cm,height=5cm]{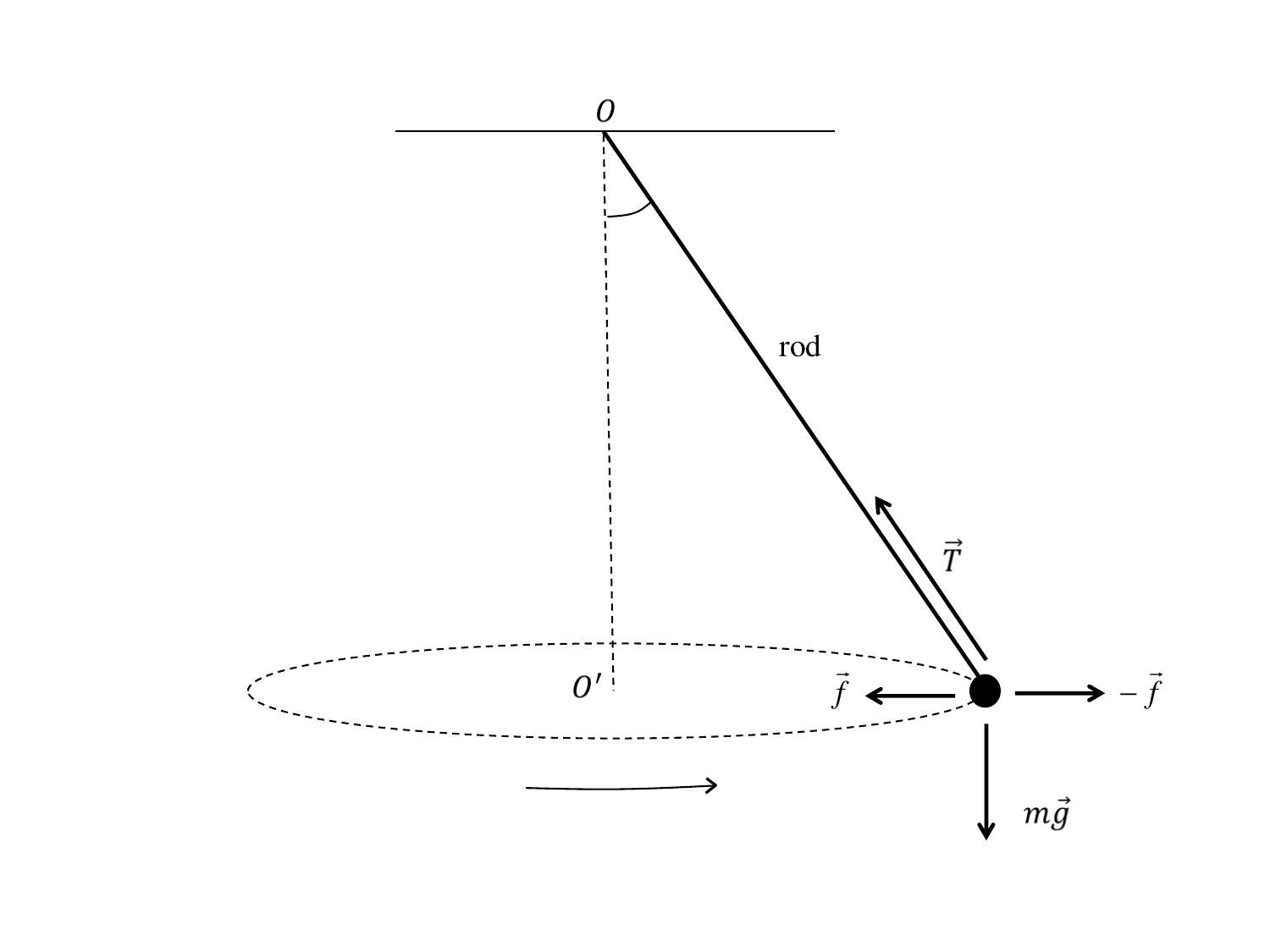}
\caption[Fig1]{\label{fig1} A rotating object composed of point-mass
with mass $m$ and a massless rod. 
}
\end{figure}
Let us consider a model, an object composed of a point-mass with
mass $m$ and a rod in Fig.1. We assume that the mass-point rotates
with a fixed angular speed making a circle drawn with dotted line.
For simplicity of discussion, we assume that the rod has negligible
mass. The rod has a joint a origin $O$. That is, its upper end is
fixed at $O$ but it can freely rotate.  We assume that the
hinge can give no torque on the rod. Here the origin with which the
torque is defined is $O$. Thus force by the rod, denoted by
$\vec{T}$, should be along the rod, like in the case of a string.

Now let us consider the same question that arises for the rotating
object: Why does not the object move downward by gravity? This
question can be answered in two standard ways.

{\bf Explanation of type 1 for rotating object}: {\it Newton's laws explain the motion: Force on
the point-mass, $m\vec{g}+\vec{T}= \vec{f}$, namely sum of gravity $m\vec{g}$
and tension $\vec{T}$, accelerates the point-mass inwardly such that
it makes a circular motion \cite{Hal93}.}

{\bf Explanation of type 2 for rotating object}: {\it We adopt a non-inertial frame co-moving with the point-mass. Fictitious force in the non-inertial frame is
centrifugal force denoted by $-\vec{f}$. The $-\vec{f}$
balances two other forces, that is, $m\vec{g}+\vec{T}-\vec{f}=0$.
Hence the point-mass has zero acceleration in the non-inertial
frame.}

Let us give other explanations. Here we adopt an inertial frame, and we focus on ``the rod" not on the object composed of the rod and the point-mass. Then we show that torque exerted on the rod is zero
\footnote{Exactly speaking, mass of the rod is non-zero. Thus torque
on the rod in vertical direction should be non-zero. In this case,
however, analysis is complicated. For simplicity, we assume that rod
has zero mass.}. If the torque is not zero, the rod will turn making the point-mass
move upward or downward. Now let us estimate torque on the rod. In
order to do that, we need to know forces which act on the rod. There
are two forces on the rod. One is force by hinge. However, the force
by hinge gives zero torque because it acts on the origin $O$. The
other is force by the point-mass. Our task is to estimate force.

{\bf Proposition-1}: {\it Here $\vec{f}$ is force that is required to maintain the (rotational) motion of the point-mass. Let force on the rod by the point-mass denoted by $\vec{U}$. Then we have $\vec{U}=-\vec{f}+m\vec{g}$.}

Let us see why Proposition-1 holds. Let us assume in Fig.1 that the
gravity is removed while the motion of object is maintained somehow.
Force that should be given on the point-mass to maintain the
constant rotation is $\vec{f}$ and the only thing that can exert
force on the point-mass is the rod. Thus it should be that the rod
is giving a force $\vec{f}$ to the point-mass. By the third law of
Newton, the force on the rod by the point-mass is $-\vec{f}$.
Intuitive interpretation of the force $-\vec{f}$ will be given later.
Now let us assume that the gravity is restored. There can be no
change in the force on the point-mass since the motion of point-mass
is the same. However, the force is differently composed. In case of
zero gravity, the force on the point-mass is solely provided by the
that of the rod. In case of non-zero gravity, the force on the
point-mass is sum of that by the rod and gravity on the point-mass,
$m\vec{g}$. Therefore, the force by the rod must be $\vec{f}
-m\vec{g}$. However, due to the third law of Newton, the force by
the point-mass on the rod is an opposite of the force. That is,
force $\vec{U}$ on the rod  by the point-mass is $-\vec{f}+m\vec{g}$.
$\Box$

It is notable that the Proposition-1 applies to not only the motion
in Fig. 1 but to motion of spinning top in Fig. 2.

Using Proposition-1, we get,

{\bf Explanation of type 3 for rotating object}: {\it The force on the rod by the point-mass, $-\vec{f}+m\vec{g}$, gives zero torque on the rod. Thus the rod will not make turning motion that makes the point-mass move vertically.}

Let us see what will happen if only the force $-\vec{f}$ were exerted
on the rod. In this case, the rod gets a torque that turns the rod
so that the point-mass moves upward. In this sense, the rotation of
point-mass gives a ``floating force". Similarly, if only gravity
$m\vec{g}$ were exerted, the rod get an opposite torque that turns
the rod so that the point-mass moves downward. We get an explanation
of Barker type \cite{Bar60}.

{\bf Explanation of type 4 for rotating object}: {\it The gravity $m\vec{g}$ gives a torque to turn the rod such that the point-mass moves downward. However, the force $-\vec{f}$ gives a torque to turn the rod oppositely such that the point-mass moves upward. The two torques are balanced and thus the mass-point does not move vertically.}

\subsection{Step 2: A spinning top}
Our arguments here are mostly in parallel with those in the previous subsection. The difference is that we deal with a spinning top as in Fig. 2: The top is a simplified one composed of four
point-masses with mass $m$ and a massless frame. Initially, when
$t=0$, four point-masses are in the $x-y$ plane, the top spins about
$z$ axis with angular speed $\omega$, and the top precesses about
$y$ axis with angular speed $\Omega$. The frame is
composed of a stem with length $R$ and four branches with length
$r$. Initial speed $v$ of the point-mass $a$ is given by $v= r
\omega + R \Omega$.
\begin{figure}[htbp]
\includegraphics[width=8cm,height=6cm]{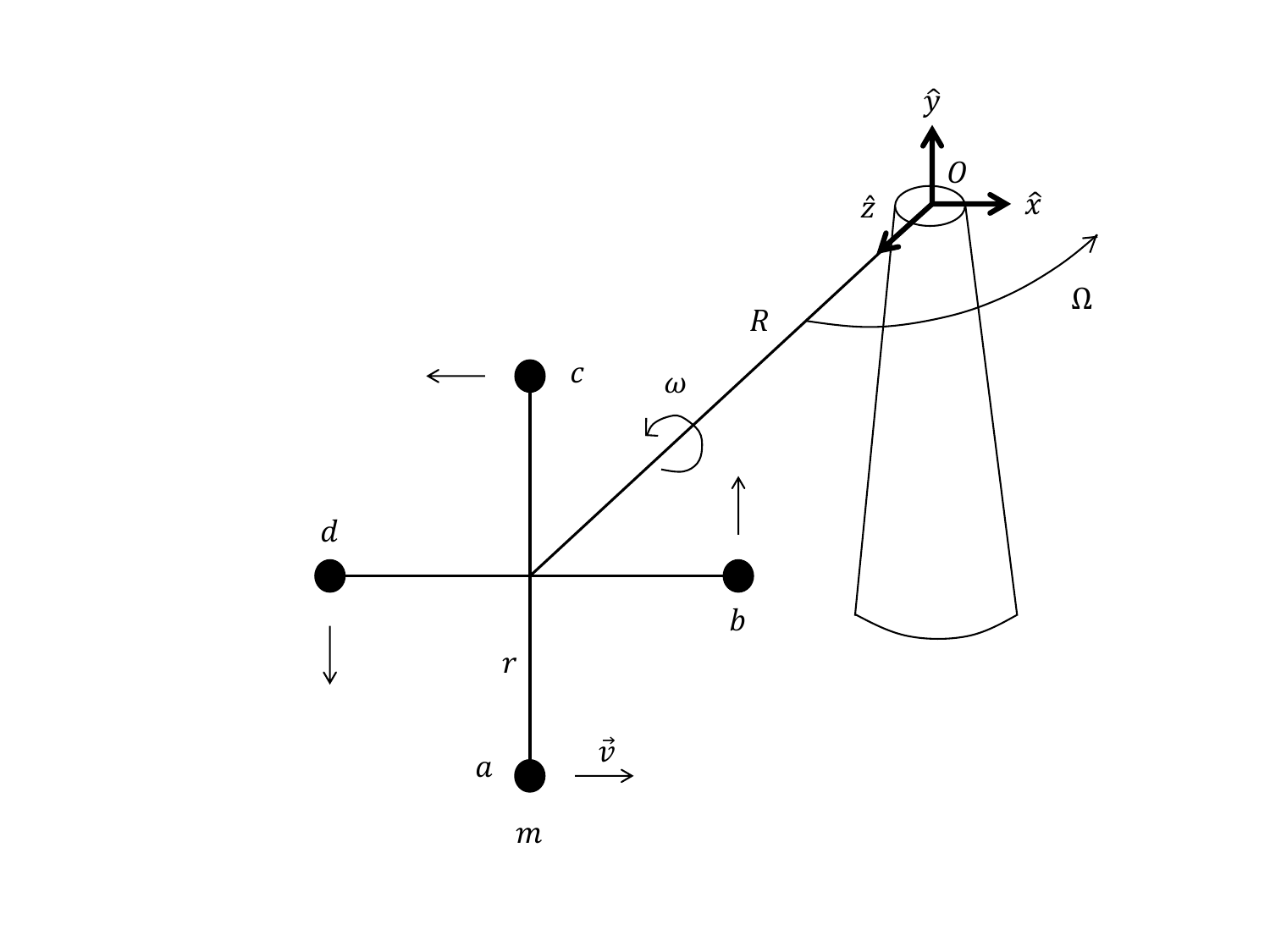}
\caption[Fig1]{\label{fig2} A simplified top composed of four
point-masses of mass $m$ and a massless frame. The frame is composed
of a stem with length $R$ and four branches with length $r$.
Initially, four point-masses are in the $x-y$ plane. The top spins
about $z$ axis with angular speed $\omega$. The top precesses about
$y$ axis with angular speed $\Omega$. Initial speed $v$ of the
point-mass $a$ is given by $v=  r \omega + R \Omega$.}
\end{figure}

Now let us calculate torque on the frame with respect to origin $O$. We will apply the Proposition-1. What we need is the force
$\vec{f}$. Force $\vec{f}$ on the point-mass by the frame can be calculated from trajectory that the point-mass makes.

Let us calculate trajectory of each point-mass in order to get
acceleration of it. Let us set up coordinate systems as in Fig. 3,
where unit vectors $\hat{\bf x},\hat{\bf y},\hat{\bf z}$ and $
\hat{\bf x}^\prime, \hat{\bf y}^\prime, \hat{\bf z}^\prime$ are
instantaneous rectangular coordinate systems co-moving with the
stem.
\begin{figure}[htbp]
\includegraphics[width=8cm,height=6cm]{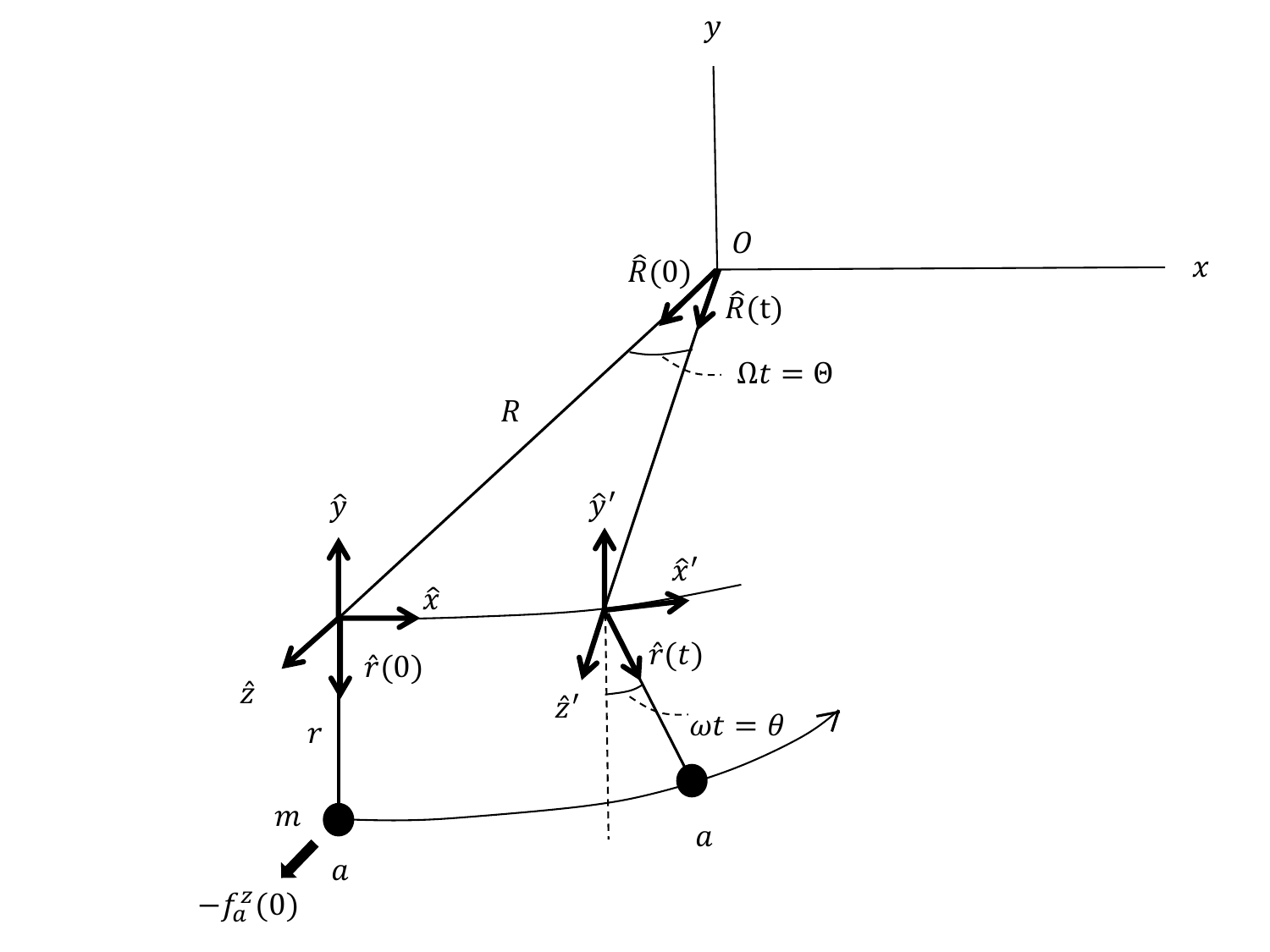}
\caption[Fig1]{\label{fig3} (Short-cut Explanation) A trajectory that the point-mass $a$
makes. The trajectory is curved toward origin $O$ as well as toward $y$ direction. Thus force on the point-mass $a$ along the
$z$ direction is non-zero. Reaction of the point-mass on the branch
along the $z$ direction on the branch, $-f_a^z(0)$, is an
opposite of the force. The other end of the stem is fixed
at the origin $O$ and the frame composed of the stem and four
branches is a solid body. Thus, the frame would turn such that the
point-mass moves upward, if the torque by the reaction
$-f_a^z(0)$ were the only one exerted on the frame. Here unit
vectors ${\hat{\bf x},\hat{\bf y},\hat{\bf z}}$ and $\hat{\bf x}^\prime,
\hat{\bf y}^\prime, \hat{\bf z}^\prime$ are instantaneous rectangular
frames co-moving with the stem.}
\end{figure}
The relationship between the two coordinate systems is given by
\begin{equation}
\label{A} \hat{\bf y}^\prime= \hat{\bf y}, \hspace{5mm}
          \hat{\bf x}^\prime= \cos \Theta \cdot \hat{\bf x}- \sin \Theta
          \cdot \hat{\bf z},
\end{equation}
as we see.

Using the coordinate system we calculate trajectory of each
point-mass. First, we calculate trajectory of point-mass $a$, $\vec{L}_a(t)$. The trajectory is sum of a location vector $\vec{R}(t)= R \hspace{1mm} \hat{\bf R}(t)$ and another location vector $\vec{r}(t)= r \hspace{1mm}\hat{{\bf r}}(t)$,
\begin{eqnarray}
\label{B} \vec{L}_a(t)= \vec{R}(t)+ \vec{r}(t),
\end{eqnarray}
where $\hat{{\bf R}}(t)$ and $\hat{{\bf r}}(t)$ are unit vectors as
shown in Fig. 3. We can see that
\begin{equation}
\label{C} \vec{R}(t)= R \hspace{1mm}\hat{{\bf R}}(t)=  R \cos
(\Omega t)\hspace{1mm} \hat{{\bf z}} + R \sin (\Omega t)
\hspace{1mm} \hat{\bf x},
\end{equation}
and that
\begin{eqnarray}
\label{D} \vec{r}(t) &=& r \sin (\omega t) \hspace{1mm} \hat{\bf x}^{\prime} - r \cos (\omega t)\hspace{1mm} \hat{{\bf y}}^{\prime}
\nonumber\\
&=& r \sin (\omega t) [ (\cos \Theta)\hspace{1mm} \hat{\bf x}-
(\sin \Theta) \hspace{1mm}\hat{{\bf z}}] - r
\cos (\omega t)\hspace{1mm} \hat{{\bf y}} \nonumber\\
&=& r \sin (\omega t) (\cos  \Omega t)\hspace{1mm} \hat{\bf x}- r
\cos (\omega
t)\hspace{1mm} \hat{{\bf y}} \nonumber\\
&& - r \sin (\omega t)(\sin  \Omega t)\hspace{1mm} \hat{{\bf z}},
\end{eqnarray}
where Eq. (\ref{A}) is used. By combining Eqs. (\ref{B})-(\ref{D}),
we obtain
\begin{eqnarray}
\label{E} \vec{L}_a(t)
&=& [R \sin (\Omega t)+ r \sin (\omega t) (\cos \Omega t)]
\hspace{1mm} \hat{\bf x}\nonumber\\
&& - r \cos (\omega t)\hspace{1mm} \hat{{\bf y}}\nonumber\\
&& +[R \cos (\Omega t) - r \sin (\omega t)(\sin \Omega t)]
\hspace{1mm} \hat{{\bf z}}.
\end{eqnarray}
From Eq. (\ref{E}), we obtain
\begin{eqnarray}
\label{F} \frac{d^2\vec{L}_a(t)}{d t^2} &=& [-R \Omega^2 (\sin
\Omega t) - r \omega^2 \sin(\omega t) \cos(\Omega t)\nonumber\\
&& - 2 r \omega \Omega \cos(\omega t)\sin (\Omega t) -r \Omega^2
\sin(\omega t) \cos(\Omega t)]
\hspace{1mm} \hat{\bf x}\nonumber\\
&& + r \omega^2 \cos (\omega t) \hspace{1mm} \hat{{\bf y}}\nonumber\\
&& +[-R \Omega^2 \cos(\Omega t)+ r\omega^2 \sin(\omega t)
\sin(\Omega t) \nonumber\\
&& - 2 r \omega \Omega \cos(\omega t) \cos(\Omega t) \nonumber\\
&& + r \Omega^2 \sin(\omega t) \sin(\Omega t) ] \hspace{1mm}
\hat{{\bf z}}.
\end{eqnarray}
What we consider is the situation when $t=0$. From Eq. (\ref{F}), we
get
\begin{eqnarray}
\label{G} \frac{d^2\vec{L}_a(0)}{d t^2} = r \omega^2 \hspace{1mm}
\hat{{\bf y}} +[-R \Omega^2 -2 r \omega \Omega ] \hspace{1mm}
\hat{{\bf z}}.
\end{eqnarray}
Thus the force $\vec{f}_a(0)$ on the point-mass $a$ when $t=0$ is
\begin{eqnarray}
\label{H} \vec{f}_a(0) &=& m\frac{d^2\vec{L}_a(0)}{d t^2} \nonumber\\
&=& mr \omega^2 \hspace{1mm} \hat{\bf y} +m[-R \Omega^2 -2 r
\omega \Omega ] \hspace{1mm} \hat{\bf z}.
\end{eqnarray}
Let us give an interpretation of Eq. (\ref{H}). The first term, $mr
\omega^2 \hspace{1mm} \hat{\bf y}$, is due to the rotation
involved with angular speed $\omega$. However, the second
term, $m[-R \Omega^2 -2 r \omega \Omega ] \hspace{1mm} \hat{\bf z}$, is due to combination of both rotations, the one involved with
angular speed $\omega$ and the other one involved with angular speed
$\Omega$. The force $\vec{f}_a(0)$ along $z$ direction is non-zero, which can also be seen from trajectory shown in Fig. 3: The trajectory is curved toward origin
$O$ as well as toward $y$ direction.

Similarly, we can calculate net
forces on other point-masses when $t=0$:
\begin{eqnarray}
\label{I} \vec{f}_b(0) &=& m\frac{d^2\vec{L}_b(0)}{d t^2}
\nonumber\\
 &=& m[-r \omega^2-r \Omega^2] \hspace{1mm} \hat{\bf x} +m[-R
\Omega^2] \hspace{1mm} \hat{\bf z},
\end{eqnarray}
\begin{eqnarray}
\label{J} \vec{f}_c(0) &=& m\frac{d^2\vec{L}_c(0)}{d t^2} \nonumber\\
&=& -mr \omega^2 \hspace{1mm} \hat{\bf y} +m[-R \Omega^2 + 2 r
\omega \Omega ] \hspace{1mm} \hat{\bf z},
\end{eqnarray}
and
\begin{eqnarray}
\label{K} \vec{f}_d(0) &=& m\frac{d^2\vec{L}_d(0)}{d t^2}
\nonumber\\
 &=& m[r \omega^2+r \Omega^2] \hspace{1mm} \hat{\bf x} +m[-R
\Omega^2] \hspace{1mm} \hat{\bf z}.
\end{eqnarray}
Note that $ \vec{f}_a(0)$ and $ \vec{f}_b(0)$ are similar to  $
\vec{f}_c(0)$ and $ \vec{f}_d(0)$, respectively. Reaction
on branch by each point-mass is $-\vec{f}_i(0)$ where $i=a,b,c,d$. By Proposition-1, force on the frame by
each point-mass, $\vec{U}_i$ is given by
\begin{equation}
\label{L-2} \vec{U}_i= -\vec{f}_i(0)+m \vec{g}.
\end{equation}

Now we are prepared to calculate torque $\vec{\tau}_i$ at $t=0$, due
to force $\vec{U}_i$. First let us separately calculate sum of
torques due to term $m \vec{g}$ of four
point-masses,
\begin{equation}
\label{L-3} \vec{\tau}^g= 4mgR\hspace{1mm} {\hat{\bf x}},
\end{equation}
where ${\bf \hat{x}}$ is the unit vector in $x$-direction. Let us
calculate remaining torque $\vec{\tau}_i^f$ due to term
$-\vec{f}_i(0)$: By inspecting spatial arrangement of vectors in Eqs. (\ref{I})and (\ref{K}), we can see that
$\vec{\tau}_b^f$ and $\vec{\tau}_d^f$ involved with point-masses $b$
and $d$ cancel each other, $\vec{\tau}_b^f+\vec{\tau}_d^f =0$.
However, by inspecting spatial arrangement of vectors in Eqs.
(\ref{H}) and (\ref{J}), we can see that sum of torques
involved with point-masses $a$ and $c$ is non-zero, that is,
$\vec{\tau}_a^f + \vec{\tau}_c^f = -4m r^2 \omega \Omega {
\hat{\bf x}} $. Therefore,
\begin{equation}
\label{M} \vec{\tau}^f= \sum_i \vec{\tau}_i^f= - 4m r^2 \omega
\Omega {\hat{\bf x}}.
\end{equation}
By Eqs. (\ref{L-3}) and (\ref{M}), we get total torque,
\begin{equation}
\label{N} \vec{\tau}= \vec{\tau}^g+ \vec{\tau}^f=(4mgR\hspace{1mm}- 4m r^2 \omega \Omega) {
\hat{\bf x}}.
\end{equation}
However, if a condition,
\begin{equation}
\label{O}  gR= r^2 \omega \Omega
\end{equation}
is satisfied, the total torque is zero. This means that frame does
not make turning motion which will make the point-masses vertically.
Interestingly, {\it this is the condition for the precession of the
top found in text books \cite{Hal93}.}

{\bf Explanation of type 3 for spinning top}: {\it The total torque on the frame is zero.
Thus the frame does not make turning motion that will make the
point-masses move vertically. Namely the frame doesn't fall.}

However, the other end of frame is fixed at the origin $O$ and the
frame is a solid body. Thus, if the torque $\vec{\tau}^g$ and
$\vec{\tau}^h$ were separately exerted on the frame, the frame
would turn such that the point-masses move downward and upward,
respectively. Now we can give a Barker-type explanation.

{\bf Explanation of type 4 for spinning top}: {\it The downward force on the frame due to the
torque by gravity, $\vec{\tau}^g$, is balanced by the floating
(upward) force due to the torque by point-masses, $\vec{\tau}^h$.
Thus spinning top does not collapse.}

\subsection{Other motions of a top}
As discussed in Ref. \cite{Bar60}, it is interesting to see
that the explanations can be applied to other motions of the top, the
nutation \cite{Fey63}. Contrary to our simple-minded notion, even a spinning top does fall depending on situation. Let us consider a motion described in Ref. \cite{Fey63}. ``If we were to hold the axis absolutely fixed, so
that it cannot process in any manner (but the top is spinning) then
there is no torque acting, not even a torque from gravity, because
it is balanced by our fingers. But if we suddenly let go, then there
will instantaneously be a torque from gravity. Anyone in his right
mind would think that top would fall, and that is what it starts
to do, as can be seen if the top is not spinning too fast. The gyro
actually does fall, as we would expect. ..." How can we explain the
``falling of spinning top with (temporarily) fixed axis"? We can find an explanation of the Explanation of type 3.

{\bf Explanation of type 3 for the case when top falls}: {\it Because the spinning axis is (temporarily)
fixed, as we can see, there is no floating force
obtained by precession of the axis. Therefore, the top falls down
due to downward force by gravity.}

However, the falling of spinning top with (temporarily) fixed axis
does not last so long. The top would begin to move horizontally such
that the top makes a cycloid, as described in Fig. 20-5 of Ref.
\cite{Fey63}. We can also give an explanation for this motion.

{\bf Explanation of type 3 for horizontal acceleration of top}: {\it The falling makes the top ``precess'' toward
$-y$ direction temporarily. However, as we have seen in section II,
a precession toward $+x$ direction induces a force toward $-y$
direction on the frame. By the same mechanism, the precession toward
$-y$ direction induce a force toward $+x$ direction on the frame.
Therefore, the falling top accelerates in $+x$ direction also.}

Now we can understand why the top makes a cycloid at least qualitatively.
\section{Discussion about Eastman's explanation}
We briefly review the Eastman's model \cite{Eas75,Edw77} in Fig. 4.
\begin{figure}[htbp]
\includegraphics[width=8cm,height=6cm]{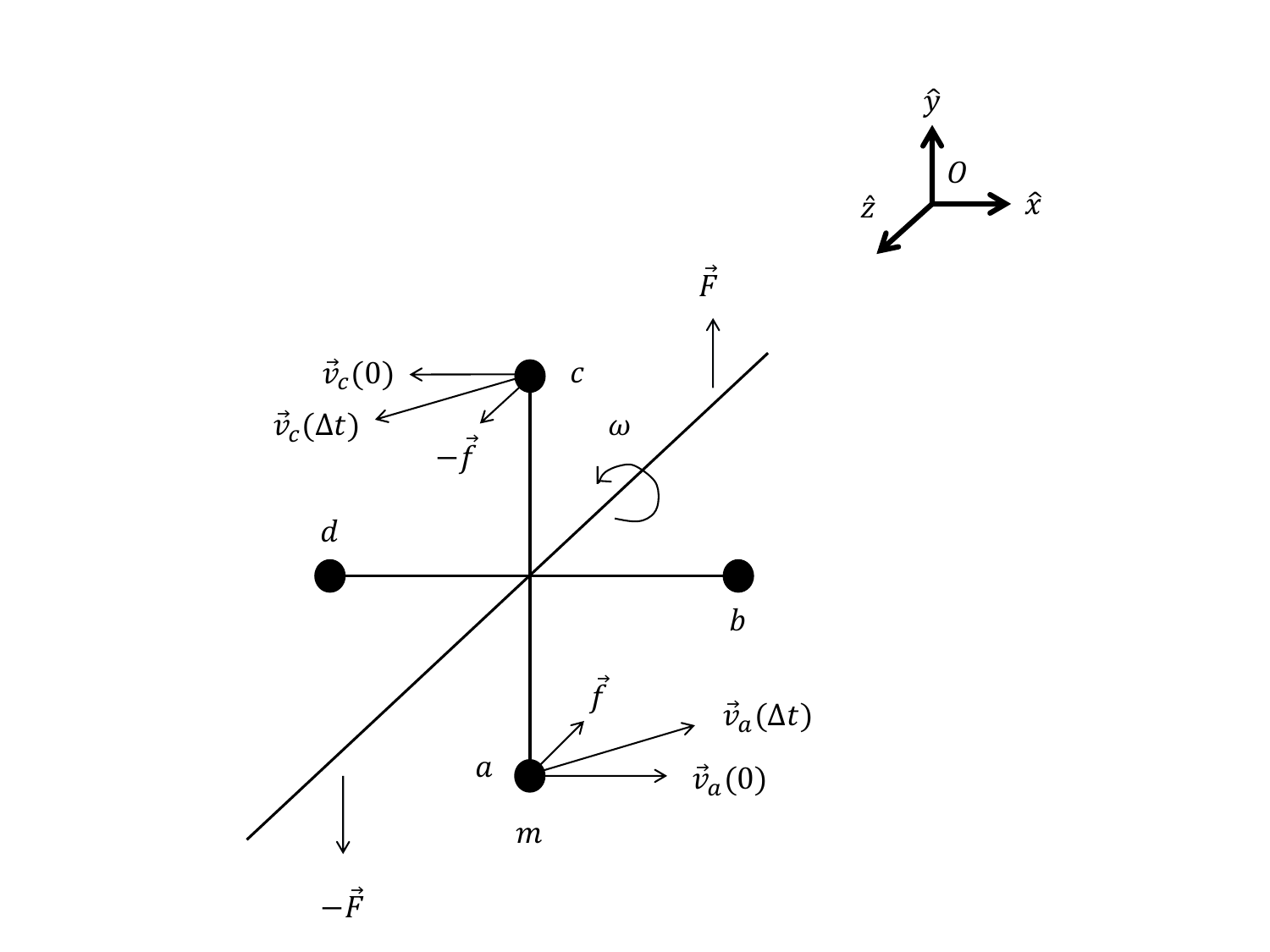}
\caption[Fig1]{\label{fig2} The Eastman's model \cite{Eas75,Edw77}.
Assume that impulses, $\vec{F}$ and $-\vec{F}$, are applied during a
time duration $\Delta t$, to a spinning top. These impulses induces
forces along the $z$ axis, $\vec{f}$ and $-\vec{f}$, on the point-masses $a$ and
$c$, respectively. The forces, $\vec{f}$ and $-\vec{f}$, then change velocities
$\vec{v}_a(0)$ and $\vec{v}_c(0)$ of point-masses $a$ and $c$ to $\vec{v}_a(\Delta t)$ and
$\vec{v}_c(\Delta t)$, respectively. Here $\vec{v}_a(\Delta t)=\vec{v}_a(0) +(\vec{f}/m) \Delta
t$ and $\vec{v}_c(\Delta t)=\vec{v}_c(0)- (\vec{f}/m) \Delta t$. Here $m$
is mass of each point-mass. As a result, the spinning axis of the
top is rotated by an angle that $\vec{v}_a(0)$ and $\vec{v}_a(\Delta t)$ make.
That is, the axis of the top precesses in a direction that is
perpendicular to the impulses.}
\end{figure}
Assume that impulses, $\vec{F}$ and $-\vec{F}$, are applied during a time duration $\Delta t$, to a spinning top. These impulses induces
forces along the $z$ axis, $\vec{f}$ and $-\vec{f}$, on the point-masses $a$ and $c$  along $z$ and $-z$ direction, respectively. The forces, $\vec{f}$ and $-\vec{f}$, then change velocities
$\vec{v}_a(0)$ and $\vec{v}_c(0)$ of point-masses $a$ and $c$ to $\vec{v}_a(\Delta t)$ and $\vec{v}_c(\Delta t)$, respectively. Here $\vec{v}_a(\Delta t)=\vec{v}_a(0) +(\vec{f}/m) \Delta
t$ and $\vec{v}_c(\Delta t)=\vec{v}_c(0)- (\vec{f}/m) \Delta t$.
Here $m$ is mass of each point-mass. As a result, the spinning axis of the top is rotated by an angle that $\vec{v}_a(0)$ and $\vec{v}_a(\Delta t)$ make. The axis of the top precesses in a direction that is perpendicular to the impulses.

Eastman's explanation has an advantage of being simple. However,it also has difficulties.  First, it is not clear how the impulses, $\vec{F}$ and $-\vec{F}$,
induces forces, $\vec{f}$ and $-\vec{f}$. What the authors of
had in mind seems to be the following.  The impulses
make point-masses $a$ and $c$ move slightly along $z$ direction. Then the motion induces the forces $\vec{f}$ and $-\vec{f}$. In
contrast, the impulses do not make point-masses $b$ and $d$
move and thus no forces are induced on the point-masses $b$ and
$d$. However, it is not clear how the motions can induce the forces $\vec{f}$ and $-\vec{f}$ in the same direction.

Second, actual trajectories of point-masses are not taken into
account in the Eastman's model. The Eastman's
model cannot explain the fact that the same spinning top either does
or does not fall down depending on initial condition. According to the Eastman's model the direction of precession is unchanged as long as the direction of impulses are the same. However, the impulses
provided by the gravity is unchanged. This implies that the spinning top would not fall down, which contradicts facts.

\section{Conclusion}
Barker's explanation about why spinning top does not collapse got to the heart of the problem: Reaction of top's point-masses exert torque on top's frame, that balances torque due to gravity. Here we supplemented the explanation. More rigorous conceptual-framework and calculations for the explanation are given. Our presentation may work as a scaffolding for the explanation. Another motion of top, the nutation, can also be understood in terms of the intuitive explanation. We discussed another intuitive explanation.
\acknowledgments
I am grateful to Profs. Intaek LIM and Jongwon PARK for helpful discussions. This study was supported by Basic Science Research Program through the National Research Foundation of Korea (NRF) funded by the Ministry of Education, Science and Technology (2010-0007208).


\end{document}